# An Argument for the Non-Arbitrary Existence Hypothesis using Anthropic Reasoning

Toby Pereira

## Abstract

This paper uses anthropic reasoning to argue for the Non-Arbitrary Existence Hypothesis (NAEH). Nick Bostrom's Self-Sampling Assumption (SSA) combined with NAEH is compared against SSA without such an assumption and also SSA with the Self-Indication Assumption (SIA). When considered in the light of various thought experiments, including the Incubator Gedanken, the Doomsday Argument, the Presumptuous Philosopher and Bostrom's Adam and Eve thought experiments, SSA with NAEH is found to give the most plausible results. The possibility of principles not explicitly considered in the paper being superior also is discussed but found to have limited plausibility.

## 1. Introduction

The question of why there is something rather than nothing is considered one of the greatest questions of philosophy, but given that there is something, we would also want to know why things are this way rather than some other way. The Non-Arbitrary Existence Hypothesis (NAEH) answers this by saying that things can only be the way they are.

**Non-Arbitrary Existence Hypothesis (NAEH) – Given that anything exists at all, things are entirely non-arbitrary and they are the only way they could possibly be.**

For example, if the universe we live in would be arbitrary when considered in isolation, then there must be multiple, perhaps even an infinite number of, universes. Indeed, you might wonder why this particular universe exists, with its particular laws of physics and its particular state of matter. NAEH removes any arbitrariness. If another universe is equally plausible, then according to NAEH, it too must exist. NAEH answers the question of why things are this way rather than that by saying that things are indeed that way elsewhere, as long as that way is equally plausible. Of course, the existence of this universe in isolation is not ruled out by NAEH, unless it can be proven that such a state of affairs would be arbitrary.

For a simple example, if you are in a situation where two equally plausible things could have happened (perhaps such as a coin landing on heads or tails), you will only experience one of these in your part of reality. But according to NAEH, there would be another part of reality where the other outcome is realised.

Plausibility here is not subjective plausibility – how plausible a reality may seem to an observer – but objective plausibility, based on all of the relevant facts (whether physical, metaphysical, mathematical, logical etc.). It might be that in certain specific heads-landing scenarios, there is no corresponding tails-landing scenario because of asymmetries between the two. However, for the examples in this paper, we will assume that they are equally plausible. We could, in any case, devise other ways of implementing a 50/50 chance that do not involve anything as crude as tossing a coin, but for simplicity coin tosses are considered standard for 50/50 situations.

A detailed discussion of the mechanism of how NAEH might be realised is beyond the scope of this paper, but see the discussion in Pereira (2014), and also Tegmark (e.g. 2014) where he discusses his Mathematical Universe Hypothesis, which simply stated says "Our external physical reality is a mathematical structure" (p. 254), and that all mathematical structures are an external reality. Also David Lewis's modal realism (e.g. 1986) is the view that all possible worlds are as real as this one. There are potentially many ways that NAEH can be realised. This paper gives a more indirect argument for NAEH by showing that it gives more reasonable results to anthropic conundrums than do other principles.

With the aid of various thought experiments, NAEH will be compared against the alternatives – an arbitrary Small World and an arbitrary Big World. In a Small World, existence is finite and small enough so that in a given scenario, the chances of its replication elsewhere is negligible. For example, if a scenario involves the toss of a coin, it can be assumed that it will land on only one of heads or tails, and there will not be an identical, or virtually identical, scenario elsewhere with the opposite result, which could interfere with an individual's anthropic reasoning processes. A Big World can be infinite in size, but need not be, and can consist of many universes. If finite, it would be big enough so that a scenario could have many identical or near-identical duplicates elsewhere.

The thought experiments and anthropic tools used in this paper largely come from Bostrom (2002), which is arguably the most comprehensive discussion of anthropic reasoning. We will start with Bostrom's Self-Sampling Assumption principle:

> *Self-Sampling Assumption*:
>
> (SSA) One should reason as if one were a random sample from the set of all observers in one's reference class. (p. 57)

Bostrom discusses who or what should be considered to be in the reference class (e.g. whether a chimpanzee would count in the same class as a human) (pp. 69-72).

Bostrom argues that SSA or at least something like it is needed to account for the Freak Observer Problem (pp. 51-57). According to our observations, the cosmic microwave background temperature is 2.7 Kelvin. However, due to random fluctuations, a tiny proportion of observers in the Universe might observe it to be e.g. 3.1 Kelvin (especially if there are a lot of observers and the universe is Big). For us to be confident in our own observations and that we are not making a freak observation, we must use a probabilistic assumption for our own identity – something like SSA.

Bostrom does actually extend SSA further to take into account individual observer moments, rather than observers individually over their whole life:

> *Strong Self-Sampling Assumption*:
>
> (*SSSA)* One should reason as if one's present observer-moment were a random sample from the set of all observer moments in its reference class. (p. 162, italics in original)

This was later extended even further so that observers are weighted by the size of their minds in terms of how much consciousness they produce (Pereira, 2017):

> Super-Strong Self-Sampling Assumption (SSSSA) – One should reason as if one's present observer-moment were a random sample from the set of all observer-moments in its reference class, with the probabilities weighted proportionally according to observer-moments' size in consciousness-spacetime. (pp. 4-5)

However, whether to use SSA, SSSA, SSSSA, or some other similar principle would be orthogonal to the discussion of the principles being compared in this paper, so for simplicity, SSA will be assumed in this paper.

SSA will be considered with and without Bostrom's Self-Indication Assumption:

> (SIA) Given the fact that you exist, you should (other things equal) favor hypotheses according to which many observers exist over hypotheses on which fewer observers exist (p. 66)

Specifically (although not explicitly mentioned by Bostrom in the definition) this favouring should be proportional to the number of observers.

Bostrom is not a big fan of SIA, but thinks it is at least worthy of consideration:

> SIA may seem *prima facie* implausible, and we shall argue in chapter 7 that it is no less implausible *ultimo facie*. Yet some of the more profound criticisms of specific anthropic inferences rely implicitly on SIA. In particular, adopting SIA annihilates the Doomsday argument. (p. 66, italics in original)

We will get to the Doomsday argument in due course, of course.

There is a certain intuitive appeal to SIA. The basic thinking is that you might not have existed. You might consider yourself "lucky" to exist. If there was only one person and that person happened to be you, you would be extremely lucky. If there were two people and one of them happened to be you, you would still be very lucky, but only half as lucky, and so on. I will not attempt to argue for or against this position, but just present it here as a potential source of intuition that could lead one to conclude that SIA is reasonable.

In some of Bostrom's examples, he asks us to consider various set-ups that exist alone in the world (or indeed the universe and beyond) so that there can be no other similar observers in the relevant reference class outside the experimental set-up existing by chance. These experiments rely on a Small World. However, NAEH does not actually permit this, since Bostrom's scenarios are arbitrary. So when considering the examples with NAEH, all the possible outcomes must be considered as actual. For example, a Small World allows us to say that a coin lands on tails and only tails, whereas NAEH does not – according to NAEH there must be another otherwise identical experiment somewhere in existence where the coin lands the opposite way.

It may seem a cheat to deviate from the explicit premises of the thought experiment being discussed. However, we can avoid breaking the rules by using NAEH as a working anthropic principle without assuming its truth. As an analogy, when Bostrom defined SSA, he was careful not to say that we should take as a fact that we have been somehow picked at random from a sample of observers, but that we should reason *as if* we were. Similarly, reasoning as if NAEH were true does not mean that it is or that we are assuming it is. After all, we are not actually using the results from the potentially identical thought experiments taking place elsewhere. But if NAEH does hold up well as a principle we can then revisit its reasonableness as a possible truth.

Despite Bostrom's Small World set-ups, he does not argue that our reality is a Small World. In fact he seems to favour a Big World and the idea of a multiverse when discussing the Freak Observer Problem. However, his thought experiments, including the Presumptuous Philosopher, which we will come to, rely on his scenarios not being proliferated elsewhere, so would at least require any Big World to have very specific (and arbitrary) constraints, which would make it behave as if it were a Small World in some respects.

To summarise, the principles we will be considering are:

SSA + NAEH (referred to as just NAEH)
SSA as Bostrom uses it with his specific constraints (referred to as SSA-Bostrom)
SSA + SIA (referred to as just SIA)

Why not consider NAEH with SIA? As will be seen, SIA tends to have the effect of requiring each potential outcome to be considered as if it is an actual outcome, regardless of whether it is. Since under NAEH, all possible outcomes are considered to be actual outcomes, adding SIA to it would be redundant. This is not to say that they give identical results in all cases, however. On top of this, the intuitive force of SIA making our existence less lucky is lost under NAEH where existence is likely to be Big.

## 2. Incubator Gedanken

We will now see how each principle deals with some of the anthropic examples that Nick Bostrom outlined, starting with the Incubator Gedanken:

> *Incubator*, version I
>
> *Stage (a)*: In an otherwise empty world, a machine called "the incubator" kicks into action. It starts by tossing a fair coin. If the coin falls tails then it creates one room and a man with a black beard inside it. If the coin falls heads then it creates two rooms, one with a black-bearded man and one with a white-bearded man. As the rooms are completely dark, nobody knows his beard color. Everybody who's been created is informed about all of the above. You find yourself in one of the rooms. *Question:* What should be your credence that the coin fell tails?
>
> *Stage (b)*: A little later, the lights are switched on, and you discover that you have a black beard. *Question:* What should your credence in Tails be now? (p. 64, italics in original)

According to SSA-Bostrom, the answer to (a) is ½ and the answer to (b) is ⅔. At stage (a), nothing relevant has been learnt, so the probability of tails is the same as the probability prior to the experiment, which is ½. The probabilities of each outcome at (a) are as follows:

Tails and black beard: ½;

Heads and black beard: ¼;
Heads and white beard: ¼.

However, after the lights are switched on, you have learnt that you have a black beard. This rules out heads-and-white-beard while leaving the ratio between the others the same, so the probabilities are now:

Tails: ⅔;
Heads: ⅓.

This gives us the answer for ⅔ (b).

According to SIA, the answer to (a) is ⅓ and the answer to (b) is ½. For (a), given that you have come into existence, you are twice as likely to be in the situation with twice as many people created. Only one person is created if the coin lands tails as opposed to two for heads, so there is a ⅓ chance that the coin landed tails, making the answer to (a) ⅓. The probabilities of each outcome are as follows:

Tails and black beard: ⅓;
Heads and black beard: ⅓;
Heads and white beard: ⅓.

After the lights are switched on, heads-and-white-beard is ruled out while leaving the ratio between the others the same as before, which gives us:

Tails: ½;
Heads: ½.

This gives the answer ½ for (b).

Let's now turn to the Non-Arbitrary Existence Hypothesis. Under NAEH, the coin landing one particular way up is an arbitrary situation and it is just as plausible to have an identical world where the other result happens. We therefore assume that both outcomes happen. This effectively means that three people are created, and we end up with the same answers as for SIA.

It is clear why a Small World is important in this case. If Bostrom did not make this assumption here, then there could be many other identical scenarios set up across the world or universe. Across all of these, we would expect half of the coins to land on heads and half on tails, with twice as many people created in the heads scenarios. This would just lead to SSA-Bostrom giving the same results as NAEH and SIA. To generate the conflict between the principles, it is essential that there is negligible or no chance that anything similar can happen in existence. A Big World could also work, but its specifications would have to be very precisely defined so that there would be no identical or near-identical experiments set up elsewhere that give the opposite result of the coin flip.

The Incubator Gedanken isn't necessarily the best barometer for evaluating the merits of each principle, since one could argue that either set of answers is more reasonable, and not much is at stake. One argument against SSA-Bostrom, however, is that SIA and NAEH provide a more continuous solution than does SSA-Bostrom. Imagine that for a heads landing, instead of two, some other number, x, people are created. If it lands on tails, exactly one person is created, as in the original example. Let's also consider only question (a) for simplicity – what should be your credence that the coin fell tails before the lights are turned on? Under SIA and NAEH, your credence that the coin landed on tails varies continuously with x. The answer would be $^{1}/_{(x+1)}$. Higher x means means lower credence. However, under SSA-Bostrom, your credence should always be exactly ½ except when x is 0, where your credence becomes 1 – it must have landed on tails for you to exist. This discontinuity might not be fatal, but it is a bit jarring.

In any case, this is a good introductory thought experiment to highlight the mechanisms of these principles and where they differ.

## 3. Doomsday Argument

Next up is the Doomsday Argument (see p. 89 onwards). The Doomsday Argument is an argument about the number of humans who will ever exist. If we take as our reference class the set of all humans who will ever have existed, then according to SSA, each of us should act as if we are a random sample from that. As discussed, there may be some debate as to what should be included in the reference class – e.g. whether to include chimpanzees etc. – but being anything other than human has already been ruled out by observation.

Bostrom uses the figure of 60 billion for the number of humans that have existed so far. This was in 2002, but it's not worth worrying about a few billion here and there for our purposes. So we know that the total number of humans who will ever exist is a number from 60 billion upwards. But if we are a random sample, then we would not expect to be at either extreme end of this *a priori*. With no other information about humanity's likely future survival, we could come up with a 95% confidence interval for the total number of humans ever to exist – with us avoiding the earliest and latest 2.5%. This would give us a range of approximately 61.5 billion to 2400 billion, or 2.4 trillion. Or between 1.5 billion and 2.34 trillion more people.

Of course, this fails to take into account any information that we might already have that makes humans more or less vulnerable to extinction than we might take from these numbers, and we'd have to use all our knowledge in a Bayesian framework to come up with a more accurate confidence interval. Also, you know about people born after you, which changes things (because you already know that there are more than 60 billion), but it's fairly small as a proportion of the total humans so it wouldn't change too much.

In any case, given no extra information about the number of humans that are likely to exist, it does seem on the face of it that SSA-Bostrom can be used to make a prediction that limits the likely number of humans in the future, although it says nothing about why this might be the case – whether due to extinction or because we evolve into post-humans etc. There are many arguments against the Doomsday argument that are beyond the scope of this paper, but the aim here is to consider the predictions of the various principles given that we accept its basic premises.

SIA exactly cancels out the Doomsday Argument, assuming a Small World at least. According to SIA, you should proportionally favour hypotheses where more people will exist. All other things being equal, you should give the hypothesis that there are two people in total twice the credence that you are the only person, and generally the hypothesis that there are x people would be given x times the credence that you are the only person.

So, while the Doomsday Argument with SSA-Bostrom predicts that the probability of there ever being x people decreases as x increases, SIA cancels this out because more people existing makes your existence more likely. See Bostrom's work for a more rigorous mathematical explanation, but here is an example to make it clearer:

Hypothesis 1: There will have been approximately 120 billion people in total.
Hypothesis 2: There will have been approximately 1200 billion people in total.

Imagine that these were the only two possibilities and that you had equal prior credence for each (before considering any empirical data or reasoning anthropically). If you then found out that you were the 60 billionth person, then according to SSA-Bostrom, hypothesis 1 would now have a 10 in 11 chance (about 91%) of being correct, since you are 10 times as likely to be the 60 billionth person when there are 120 billion people than when there are 1200 billion people. However, according to SIA, hypothesis 2 is 10 times as likely as hypothesis 1, all else being equal, as there are 10 times as many people under hypothesis 2, so when combined with SSA, this exactly cancels out the conclusion from SSA-Bostrom.

However, unlike the Incubator Gedanken, which is a purely hypothetical scenario, the Doomsday Argument can be used as a prediction about the real world that we live in, so the premise "In an otherwise empty world" cannot be assumed to apply. There may be other planets elsewhere in the universe with intelligent life on. In that case, there would be no need to posit that a large number of beings live on any individual planet, given that that there could instead be a large number of planets with intelligent life on. This means that even under SIA, the Doomsday Argument may still have some force. But that is an aside. If we take it as a purely hypothetical scenario in a Small World, then SSA-Bostrom and SIA differ in their predictions.

NAEH agrees with SIA in cases where there are already set probabilities (e.g. involving a coin toss) concerning the number of people in the population, such as in the Incubator Gedanken, because both "parallel worlds" would be assumed to exist under NAEH. However, NAEH has no particular bias in its philosophy towards a greater number of people in general, and so gives no defence against the Doomsday Argument. NAEH therefore gives exactly the same result as SSA-Bostrom.

It is worth pointing out that Bostrom himself does not agree with the conclusion of the Doomsday Argument despite rejecting SIA. However, this is for other reasons to be discussed later. The main conclusion to be drawn here is that in the Doomsday Argument scenario, if we reasoned using NAEH, we would not conclude any differently from if we reasoned using SSA-Bostrom.

## 4. The Presumptuous Philosopher

In the case of the Incubator Gedanken, NAEH agreed with SIA, whereas in the case of the Doomsday Argument, it agreed with SSA-Bostrom, so it already has its own distinct space in the anthropic landscape. The Incubator Gedanken is a fairly abstract thought experiment without too much riding on the result, whereas the Doomsday Argument is more likely to divide opinion. However, neither result seems particularly embarrassing for any of the principles. The Doomsday Argument does not seem obviously right or wrong from the outset. For potentially embarrassing scenarios, let's first turn to the Presumptuous Philosopher, a thought experiment designed by Bostrom to embarrass SIA.

*The Presumptuous Philosopher*

> It is the year 2100 and physicists have narrowed down the search for a theory of everything to only two remaining plausible candidate theories, $T_1$ and $T_2$ (using considerations from super-duper symmetry). According to $T_1$ the world is very, very big but finite and there are a total of a trillion trillion observers in the cosmos. According to $T_2$, the world is very, very, *very* big but finite and there are a trillion trillion trillion observers. The super-duper symmetry considerations are indifferent between these two theories. Physicists are preparing a simple experiment that will falsify one of the theories. Enter the presumptuous philosopher: "Hey guys, it is completely unnecessary for you to do the experiment, because I can already show to you that $T_2$ is about a trillion times more likely to be true than $T_1$!" (p. 124, italics in original)

SSA-Bostrom is indifferent between the two theories, whereas SIA says that we should favour the theory where there are more observers, so $T_2$, which is what the presumptuous philosopher himself is arguing for. Bostrom considers the conclusion from SSA-Bostrom to be more reasonable.

> Finally, consider the limiting case where we are comparing two hypotheses, one saying that the universe is finite (and contains finitely many observers), the other saying that the universe is infinite (and contains infinitely many observers). SIA would have you assign probability one to the latter hypothesis, assuming both hypotheses had a finite prior probability. But surely, whether the universe is finite or infinite is an open scientific question, not something that you can determine with certainty simply by leaning back in your armchair and registering the fact that you exist!
>
> For these reasons, we should reject SIA. (p. 126)

However, depending on exactly how the thought experiment is designed, the presumptuous philosopher's conclusion might not be so illogical. According to NAEH, if both types of universe are equally probabilistically plausible (objectively, not just according to our limited knowledge), then both types of universe exist anyway, in equal numbers, so there would be a trillion times as many observers living in a $T_2$-type universe. This means that we would indeed be a trillion times as likely to be in a $T_2$-type universe, in agreement with SIA and the presumptuous philosopher. Bostrom introduced this example seemingly as a demonstration of the unreasonableness of SIA, but I do not think that the conclusion is so unreasonable. Although there is disagreement between the principles here, it is is not the source of embarrassment for SIA that Bostrom intended.

However, if the thought experiment was set up slightly differently, then NAEH would agree with SSA-Bostrom, with SIA on its own and in a more compromised position. In this modified scenario, physicists have equal subjective credence for $T_1$ and $T_2$, but they also know that one (and only one) of them is actually logically impossible, rather than merely not the case in our universe. They just don't know which. We can imagine that a team of logicians is working on the problem and that the result is due soon. Reasoning by SIA, it seems as if we should still conclude that $T_2$ is a trillion times as likely as $T_1$, since we are a trillion times as likely to exist if $T_2$ is the possible scenario, even though its truth is determined by logic. SSA-Bostrom would be indifferent, as we would expect. NAEH encourages us to look at all objectively plausible scenarios. Since only one of $T_1$ and $T_2$ is objectively plausible (the other being logically impossible), it does not ask us to favour $T_2$. As with SSA-Bostrom, we would simply have to wait for the logicians to present their conclusion.

Certainly, in this modified situation, SIA seems to be imposing itself on logical truth. So set up this way, the Presumptuous Philosopher thought experiment can embarrass SIA. Regardless of Bostrom's version of the experiment, the modified version is a black mark against SIA.

This is also a case where Bostrom's use of SSA implies either a Small World or arbitrary existence. While Bostrom appears in favour of a universe of potentially infinite size and indeed a multiverse (e.g. p. 51), the Presumptuous Philosopher thought experiment relies on the assumption that the size that our universe happens to be is typical of universes containing observers – i.e. that any multiverse would contain universes largely similar to this one – or that there is no multiverse. This is because in a non-arbitrary Big World, multiple universes could exist with many different laws of physics. So even if $T_1$ is true for our universe, the possibility of $T_2$ being the case for other universes (unless in only a tiny proportion of them) would still lead us to conclude that we are overwhelmingly more likely to be in a $T_2$ universe than a $T_1$ one. The parochial laws of our universe or even our particular multiverse (if we live in one) would not apply across the totality of existence under a non-arbitrary Big World hypothesis.

## 5. Adam and Eve Thought Experiments

Bostrom later presented three Adam and Eve thought experiments which are potentially embarrassing for SSA-Bostrom. Number 1:

> *First experiment: Serpent's Advice*
>
> Eve and Adam, the first two humans, knew that if they gratified their flesh, Eve might bear a child, and if she did, they would be expelled from Eden and would go on to spawn billions of progeny that would cover the Earth with misery. One day a serpent approached the couple and spoke thus: "Pssst! If you embrace each other, then either Eve will have a child or she won't. If she has a child then you will have been among the first two out of billions of people. Your conditional probability of having such early positions in the human species given this hypothesis is extremely small. If, one the other hand, Eve doesn't become pregnant then the conditional probability, given this, of you being among the first two humans is equal to one. By Bayes' theorem, the risk that she will have a child is less than one in a billion. Go forth, indulge, and worry not about the consequences!" (p. 142, italics in original)

Number 2:

> *Second experiment: Lazy Adam*
>
> The next example effects another turn of the screw, deriving a consequence that has an even greater degree of initial counterintuitiveness:
>
> Assume as before that Adam and Eve were once the only people and that they know for certain that if they have a child they will be driven out of Eden and will have billions of descendants. But this time they have a foolproof way of generating a child, perhaps using advanced *in vitro* fertilization. Adam is tired of getting up every morning to go hunting. Together with Eve, he devises the following scheme: *They form the firm intention that unless a wounded deer limps by their cave, they will have a child*. Adam can then put his feet up and rationally expect with near certainty that a wounded deer—an easy target for his spear—will soon stroll by. (p. 143, italics in original)

Number 3:

> *Third experiment: Eve's Card Trick*
>
> One morning, Adam shuffles a deck of cards. Later that morning, Eve, having had no contact with the cards, decides to use her willpower to retroactively choose what card lies top. She decides that it shall have been the dame of spades. In order to ordain this outcome, Eve and Adam form the firm intention to have a child unless the dame of spades is top. They can then be virtually certain that when they look at the first card, they will indeed find the dame of spades. (p. 144, italics in original)

These thought experiments (which I have lumped together for one discussion) have similarities with the Doomsday Argument, except that they each give a specific mechanism as to how the population might or might not increase, sometimes quite implausibly, and they are set up to embarrass SSA-Bostrom. Bostrom does not discuss them in relation to SIA (presumably because he has dismissed it by this stage), although the result would be the same as for the Doomsday Argument. The improbability of being among the first people to exist would be exactly cancelled out by many people existing, so the experiments do not result in any embarrassment for SIA.

For NAEH, you would need to know the make-up of any other worlds for an objective probability, but from Adam and Eve's point of view in possession of limited facts, the fact that they are the first two people means that they should, *all other things being equal*, consider their situation to be fairly typical, and therefore being among the first two people as fairly probable. However, all other things are not equal; there are certain signs in all these cases that this is a very contrived or arbitrary situation, and NAEH is very sensitive to this! In possession of a few facts, it would be hard not to be aware of the immense contrivedness and arbitrariness of the entire set-up, so Adam and Eve would realise that they must already be in a very improbable situation – a situation that would not be typical across the many worlds that must exist to make their world part of a non-arbitrary whole.

If Adam and Eve consider that it is likely for a world generally to spawn a population of billions, then it actually makes no difference whether in their particular world a large population results, because they must consider people from these other worlds to be in their reference class. This is where NAEH differs from Bostrom's use of SSA. SSA-Bostrom parochially deals with just the one world, which it then "forces" into shape with unlikely events based on a reference class that can be very contrived, whereas NAEH does not allow such a contrivance as it opens up the reference class to all other similarly possible worlds. NAEH, like SIA, is immune to the ramifications of these thought experiments.

With an extra tweak to the Adam and Eve experiments, things get even worse for SSA-Bostrom, as they did for SIA with the modified version of the Presumptuous Philosopher experiment. Imagine that Adam and Eve are aware of the number pi and know it to two decimal places, but they've found out a way of working it out to five decimal places. Adam and Eve form the intention that unless the fifth decimal place of pi is 7, they will have a child. This is similar to the previous Adam and Eve examples, except that it it is an attempt by SSA-Bostrom to force a mathematical truth (similar to SIA trying to force a logical truth in the modified Presumptuous Philosopher experiment), rather than just an unlikely but possible event. By the way, the fifth decimal place of pi is 9, not 7. This clearly doesn't work on NAEH because we can equally imagine universes where other Adams and Eves do the same except with different values for the fifth decimal place of pi, and pi must be the same across these universes. 7 is an arbitrary choice. SIA is immune as it is with the other Adam and Eve thought experiments.

## 6. Bostrom’s Defence of SSA

Later in the book, Bostrom goes on to defend his use of SSA from the conclusions of the Adam and Eve thought experiments, along with the conclusions of the Doomsday Argument (see pp. 162-172). However, he accepts that none of it is conclusive, and the defence does appear to be a bit *ad hoc*. It relies on a proliferation of reference classes based on time slices, arguing that observers’ knowledge of the time they exist in is enough to distinguish between reference classes. For e.g. the Doomsday Argument, we would be in a different reference class to those living before or after us, so the number of people who lived in the past or will live in the future would have no bearing on our reasoning of the total number of people who will have lived.

I have previously argued (e.g. Pereira, 2017) that we should make the reference class as inclusive as possible, and include all forms of consciousness, not just human consciousness, and certainly not just humans at a specific time. Otherwise you end up with seemingly arbitrary cut-offs for reference classes, and it potentially enables one to divide every distinct conscious thought into a different reference class, rendering all anthropic reasoning of this sort useless. It is on this basis, along with the Super-Strong Self-Sampling Assumption (SSSSA), that I argued that a proliferation of superintelligent AI is unlikely, given that it would make our conscious minds improbably impoverished in the landscape of all consciousness. David Leslie gave a thought experiment that suggests we should not discount those living at different times from our reference class:

> Imagine an experiment planned as follows. At some point in time, three humans would each be given an emerald. Several centuries afterwards, when a completely different set of humans was alive, five thousand humans would each be given an emerald. Imagine next that you have yourself been given an emerald in the experiment. You have no knowledge, however, of whether your century is the earlier century in which just three people were to be in this situation, or in the later century in which five thousand were to be in it. Do you say to yourself that if yours were the earlier century then the five thousand people *wouldn’t be alive yet*, and that therefore you’d have no chance of being among them? On this basis, do you conclude that you might just as well bet that you lived in the earlier century?
>
> Suppose you in fact betted that you lived there. If every emerald-getter in the experiment betted in this way, there would be five thousand losers and only three winners. The sensible bet, therefore, is that yours is instead the later century of the two. (1996, p. 20, italics in original)

Bostrom himself quotes this passage in support of SSA. One difference between Leslie’s thought experiment and the Doomsday Argument is that in Leslie’s experiment, you have no idea which of the two time periods you are in, whereas you do know your time period in the Doomsday scenario, arguably allowing you to exclude people living in other time periods from your reference class in the Doomsday scenario only. However, this reasoning appears to be a little strained. Certainly, if we use the argument explicitly given by Leslie and seemingly accepted by Bostrom – that of betting – then it would apply equally well to the Doomsday Argument as it would for the emeralds. Imagine if every human who ever lived had to bet whether they were proportionally among the first $^{3}/_{5003}$ or the last $^{5000}/_{5003}$ of all humans. The way to get the most winners would be for everyone to bet that they were among the last $^{5000}/_{5003}$. This is equivalent to the scenario with the emeralds.

NAEH does not require us to define the reference class in any sort of contrived way in order to produce the sensible-looking results that it does. SSA-Bostrom does, and this is damaging. It does seem that, *prima facie*, a lot of damage has been done to SSA as Bostrom uses it.

## 7. Conclusions

These thought experiments taken together show that it is not hard to contrive situations that can make either SSA-Bostrom or SIA look bad. NAEH is immune to this sort of contrivance because under NAEH we can't force all of existence and all observers into such a contrived – or indeed, arbitrary – situation, by its very definition. To clarify, although this paper has been arguing against SSA as Bostrom uses it, the argument is not against SSA per se. As discussed, SSA, or something close to it, is an underlying assumption of the paper.

We saw with the Adam and Eve thought experiments that in contrived situations, SSA-Bostrom can be used to will very improbable events to become probable, such as a wounded deer wandering by or a particular card ending up on top of the deck.

We also saw in the variant of the Presumptuous Philosopher thought experiment, SIA attempted to force a logical truth, which perhaps seems worse than the Adam and Eve examples with SSA-Bostrom, which were used merely to predict bizarre real-world coincidences. However, SSA-Bostrom can also be used on logical and mathematical truths, as we saw with the Adam and Eve pi modification.

All this is not to say that it is impossible to contrive embarrassing scenarios for NAEH (maybe I just wasn't trying hard enough), but contrived is arbitrary, and the *Non-Arbitrary* Existence Hypothesis has natural defences against that! Also the more obviously unlikely a scenario, the more it becomes unlikely for the observers anyway, and they will be able to reason from this position. With SSA-Bostrom and SIA, we can define these contrived scenarios and just declare them to be all of existence, regardless of how arbitrary they are. We can put existence in a box, which we can't do with NAEH because the box is arbitrary. This is why it would take a different type of scenario to do any damage to NAEH.

Regardless of where we previously stood on these principles – SSA-Bostrom, SIA and NAEH – NAEH seems to give the most plausible results. Obviously I haven't considered every possible principle, but other principles that could also give plausible results would still be more arbitrary than NAEH (by definition), and by Ockham's Razor, we'd need good reason to consider them above NAEH.

This is obviously open to accusations of question-begging – saying that the best solution is the non-arbitrary one on the basis that everything else is arbitrary! However, this paper goes beyond that. It shows the consequences of allowing for arbitrary existence in anthropic scenarios. We might wonder why the universe is this way rather than that, and might have been uncomfortable with the possibility that certain things were simply arbitrary, but that is not a strong reason to reject arbitrariness out of hand. These anthropic scenarios give us concrete examples where arbitrary exclusionary existence can lead to ludicrous situations. While other arbitrary existence hypotheses may allow for ways around these embarrassing results, they would have to be carefully (arbitrarily) picked from all the possible arbitrary principles to do so. We have seen that it is not just the Small World assumption that causes problems, but that Big Worlds can too depending on how they are set up.

Working our way around these restrictions within the realms of arbitrary existence in a way to specifically give plausible results would put us two layers of arbitrariness deep, and so arguably in a worse position than Bostrom's examples that allow for the creation of any specific form of existence to the exclusion of everything else. And would it just be by luck if one of these doubly arbitrary hypotheses turned out to be the correct existence hypothesis, allowing us to avoid thought experiments and potentially real-life situations with embarrassing results? It just does not seem realistic. Far better to conclude in favour of the Non-Arbitrary Existence Hypothesis. This is unless we want to simply accept embarrassing results, such as being able to create a scenario where we can simply will a particular decimal place of pi to be any digit we like. I, for one, do not!

While concluding in favour of NAEH seems to avoid a lot of problems, one might still wonder "But what if NAEH hadn't been true?" If it hadn't, we would be left with some difficult results to deal with, so it seems a bit convenient if it does happen to be true. However, the conclusion is not that NAEH might have been true and it might not have been, but we got lucky that it turned out to be so, so we don't have to deal with horrible results. To conclude in favour of NAEH is to assert its logical necessity and that there could not be an alternative reality where it doesn't hold. And while this paper has not proven the truth of NAEH, it has shown some of the consequences of a reality where it is not the case. Imagine if someone asks: "Yes, but what would be the correct answers to the questions in the Incubator Gedanken if NAEH were not the case and the experiment comprised all of existence? Would you side with SSA or SIA?" It would be legitimate to respond that there is no sensible answer one could give if it is actually a logical impossibility. Or one could respond that we can still use NAEH as a working anthropic principle without assuming its truth, which is actually what we did at the start of the paper so that Bostrom's thought experiments could be taken at face value.

But now that we have reached this point, it can be said that it is certainly an advantage for a hypothesis to hold up to scrutiny when reasoning as if it is true, so our credence for the Non-Arbitrary Universe Hypothesis must still increase from what it was before.